\documentclass[12pt,a4paper]{article}\usepackage[]{graphicx}\usepackage[]{color}
\makeatletter
\def\maxwidth{
  \ifdim\Gin@nat@width>\linewidth
    \linewidth
  \else
    \Gin@nat@width
  \fi
}
\makeatother

\definecolor{fgcolor}{rgb}{0.345, 0.345, 0.345}

\usepackage{framed}


\definecolor{shadecolor}{rgb}{.97, .97, .97}
\definecolor{messagecolor}{rgb}{0, 0, 0}
\definecolor{warningcolor}{rgb}{1, 0, 1}
\definecolor{errorcolor}{rgb}{1, 0, 0}

\usepackage{alltt}
\usepackage[T1]{fontenc}
\usepackage[utf8]{inputenc}
\usepackage{natbib}
\usepackage{float}
\usepackage{graphicx}
\usepackage{tabularx}
\usepackage{multirow}
\usepackage{afterpage}
\usepackage{rotating}
\usepackage{pdflscape}

\usepackage[hmargin=2.5cm,vmargin=3.5cm]{geometry}

\newcommand{\thisauthor}{Kristian Koerselman}
\newcommand{\thistitle}{Why Finnish polytechnics reject top applicants}

\usepackage[usenames,dvipsnames,svgnames,table]{xcolor}
\definecolor{linkcolor}{rgb}{0.10,0.28,0.46}
\usepackage{hyperref}
\hypersetup{
pdfauthor = {\thisauthor{}},
pdftitle = {\thistitle{}},
pdfsubject = {},
colorlinks=true,
citecolor=black,
filecolor=black,
linkcolor=black,
urlcolor=linkcolor
}

\IfFileExists{upquote.sty}{\usepackage{upquote}}{}
\begin{document}

\title{\thistitle}
\author{\thisauthor\footnote{Finnish Institute for Educational Research and Jyväskylä University School of Business and Economics, P.O.\ Box 35, FI-40014 University of Jyväskylä, Finland. E-mail: kristian.w.koerselman@jyu.fi.}}
\maketitle

\begin{abstract}
I use a panel of higher education clearinghouse data to study the centralized assignment of applicants to Finnish polytechnics. I show that on a yearly basis, large numbers of top applicants unnecessarily remain unassigned to any program. There are programs which rejected applicants would find acceptable, but the assignment mechanism both discourages applicants from applying, and stops programs from admitting those who do. A mechanism which would admit each year's most eligible applicants has the potential to substantially reduce re-applications, thereby shortening the long queues into Finnish higher education.
	\\[5pt]
\vfill
\end{abstract}

\section{Introduction}

Economists have helped to create and improve the centralized application systems which assign students to schools or applicants to programs,\footnote{I will use \emph{student} and \emph{applicant}, and \emph{school}, \emph{program} and \emph{college} interchangeably in what follows.} perhaps most famously in the redesign of school application systems in New York City and Boston \citep{abdulkadiroglu2005nyc,abdulkadiroglu2005boston,abdulkadiroglu2009}.\footnote{See \citep{pathak2017} for a recent overview} In doing so, they have often taken for granted that the ultimate goal of the application system should be to assign applicants as much as possible in accordance with their preferences, under the restriction that the assignment is not illegal or otherwise unjust or inequitable. I highlight a competing policy goal which has received less attention in application system design: that the most suitable applicants to each program should be assigned to that program. 

I use a panel of Finnish higher education clearinghouse data to study the centralized assignment of applicants to Finnish polytechnic programs from the perspective of educational selection. I show that even if admission at Finnish polytechnics is highly selective in terms of the proportion of applicants admitted, it is not in terms of admitted applicants' observable quality, even when measured in terms of the actual selection criteria. Large numbers of top applicants remain unassigned, and the matriculation exam grades of admitted applicants do not greatly exceed the average across all applicants. 

Labor economists and sociologists have studied educational selection as a product of applicant-specific background factors on the one hand, and of institutional features like admission criteria on the other. I argue that in a centralized admissions system, the choice of mechanism is an additional determinant of educational selection. I show that in the Finnish polytechnic assignment, the mechanism is rejecting top applicants unnecessarily. The mechanism can thus be thought to attenuate stated selection criteria. The analysis presented in this paper furthermore suggests that the extensive queuing into Finnish higher education may be a direct effect of the mechanism's failure to admit each year's most eligible applicants.

The remainder of the paper proceeds as follows. In Section 2, I give an overview of related literature. In Section 3, I describe the institutional background of higher education admissions in Finland. I describe the data and methods in Section 4, and the empirical results in Section 5. Section 6 concludes.

\section{Background}
Schools or programs will commonly admit only a limited number of applicants each. When a program is oversubscribed, some kind of choice between applicants thus has to be made. For this purpose, applicants to each program are typically sorted into a priority ordering on the basis of eligibility criteria, and admitted accordingly. 

When applications to multiple programs are combined into a single application system, this can increase the efficiency of the assignment, among others by reducing the number of programs that end up unnecessarily undersubscribed when some of the applicants they admitted accept a seat at a different program instead. In a centralized system, there is however typically more than one reasonable way to assign applicants to programs, leaving the mechanism designer to navigate the trade-offs between conflicting policy goals.

In practice, mechanism designers have typically concentrated on trying to assign applicants as much as possible in accordance with applicants' wishes on which program to attend, subject to an elimination of priority violations. An absence of priority violations is also called stability, and implies that when two applicants apply to the same program, and the more eligible applicant is not admitted, the less eligible applicant is not admitted either. Stability comes at a cost to applicant welfare in the sense that applicants are assigned to programs they prefer less, and it is often possible to make improvements to an assignment if we are willing to relax or remove the requirement that the assignment be stable \citep[e.g.][]{AbdulkadirogluSonmez2003,kesten2010}.

Apart from an elimination of priority violations, a common restriction on the assignment mechanism is that it should be strategy-proof: that applicants should not be able to improve their assignment by strategically misrepresenting their preferences. When a mechanism is not strategy-proof, this may create inequities between applicants who differ in their knowledge of the application system, or in their willingness to game it. A lack of strategy-proofness can also be an obstacle to efficiency, for example when an applicant has an incentive ex ante not to apply to a desirable program that would in fact have accepted her ex post, for example because it turned out to be undersubscribed.

The Gale-Shapley student optimal stable mechanism, also called student-proposing deferred acceptance, has become the standard against which other assignment mechanisms are judged. It creates a stable assignment which Pareto dominates all other stable assignments in terms of applicant welfare, while also being stra\-te\-gy-proof for applicants \citep{GaleShapley1962,AbdulkadirogluSonmez2003}. 

When applicants can predict accurately which programs would be willing to admit them, they may refrain from applying to programs where their assignment probability is very low, either because they are certain not to be considered at the program itself, or because they are certain to be admitted at a more preferred program instead. Though such applicants are not truthful about their preferences in a technical sense, they may re-create the student optimal stable outcome even in the absence of a mechanism that is strategy-proof \citep[][]{fack2018}, by applying exactly to the most preferred programs that would be willing to admit them. In other application systems, applicants face large uncertainty. Though uncertainty is what incentivizes applicants to apply to an exhaustive list of programs in order of preference under a strategy-proof mechanism, under other mechanisms it can instead allow applicants to credibly reveal the cardinal strength of their preferences by applying to preferred but relatively competitive programs only if they strongly prefer them, and to safer programs otherwise \citep{abdulkadirouglu2011}.

In their treatment in the mechanism design literature, priority orderings are primarily thought to confer rights on applicants.\footnote{This is perhaps the most clear in the work of \citet{kesten2010}, who designs a mechanism which allows applicants to waive specific priority rights to the benefit of other applicants.} They are restrictions on the set of legal assignments, not maximands in and of their own. While the policy maker could try to maximize program rather than applicant welfare, this possibility has not been given much consideration. Indeed, when priorities reflect programs' genuine preferences on whom to admit, they are often treated as illegitimate. \citet{GaleShapley1962} for example write that
\begin{quote}``\emph{colleges exist for the students rather than the other way around}''\end{quote}
and \citet{AbdulkadirogluSonmez2003} that 
\begin{quote}``\emph{only the welfare of students matters in the context of school choice}.''\end{quote}
	The disregard given to the welfare of independent programs seems to have carried over to settings where admission is based on laws and regulations, and where programs do not play an active role in the admission process.\footnote{The assignment problem in which programs are actors is called the \emph{college admissions problem}, while programs are passive implementers of external priority orderings in the \emph{student placement problem} and in the \emph{school choice problem} The difference between the latter two lies in whether priorities are based on achievement scores or on background variables \citep[cf.][]{sonmezunver2011}. Though the case for taking program preferences seriously is perhaps the strongest in the student placement problem, program preferences are not necessarily irrelevant in either the school choice or in the college admissions problem either.}. In these settings, seats in programs are seen as mere ``\emph{objects to be consumed}'',\footnote{C.f.\ e.g.\ \citet[Section 4]{balinski1999}: ``\emph{the very definition of Pareto efficient depends on whether or not colleges are agents.}''} and priority orderings are treated as pure applicant rights. There are cases when this is fully reasonable, for example when priorities are determined by lottery. The priority given to one applicant over another then reflects a reality that both applicants cannot be admitted, but is not an indication that the admission of one applicant should fundamentally be preferred over the admission of another. In other cases however, priorities are not based on lottery draws, but rather on eligibility criteria that reflect arguably legitimate social preferences on who should be assigned where. Program welfare represents a kind of social welfare in this setting, and there exists a genuine conflict of interest between applicants and wider society. Should a country for example try to accept the best applicants into its medicine programs, or rather the applicants who most of all want to study medicine?\footnote{A discussion on how best to assign physicians to residencies is the source of a rare example of documented reasoning about how to handle the trade-off between applicant and program welfare \citep[footnote 9]{roth1999}. \citeauthor{roth1999} highlight that neither applicants nor programs base their stated preferences on full information. Because applicants' ex ante preferences over programs are thought to be much less noisy than programs' ex ante preferences over applicants, physician welfare should be targeted in favor of residency program welfare.}

There exist a number of earlier studies of assignment mechanisms in the context of educational selection. Part of the mechanism design literature has been devoted to settings in which the policy maker wants to encourage or guarantee a certain distribution of applicant types over programs, ensuring diversity within each program in terms of gender, socio-economic background, or ethnicity \citep[cf.\ e.g.][]{echenique2015}. Notably, \citet{dur2018} provide an example of how selection criteria are attenuated by mechanism design choices in the implementation of neighborhood priorities in Boston schools. The authors argue that while 50\% of seats are reserved for applicants from the school's neighborhood, the mechanism assigns the reserved seats to a selection of neighborhood applicants likely to be assigned to the school anyway, greatly diminishing the potential effect of the quota. 

\citet{carvalho2018} study a two-stage selection process into a pair of medical schools in Brazil. A lower cost first stage is used to screen applicants before a higher cost second stage. Applicants have to choose which school to apply to before knowing their first stage examination result. The authors use estimated preference parameters to simulate counterfactual selection procedures, among others allowing applicants to apply to both schools' second stage examinations, and allowing applicants to choose which school to apply to after learning their first stage result rather than before it. When applicants can apply to both schools, each school effectively has more applicants to choose from, and more top applicants are admitted. Similarly, when applicants learn their first stage result before applying for a second stage examination seat, they can apply to the most preferred school that would accept them into its second stage examination, and in equilibrium the programs manage to select the highest scoring feasible applicants. A combination of applicant uncertainty and limits on the number of programs applicants can effectively apply to can thus be thought to attenuate selection compared to these counterfactuals.

Even under a strategy-proof mechanism, uncertainty may attenuate selection. \citet{chen2019} for example study the Mexico City high school match. Though the mechanism is strategy-proof, applicants fail to apply to schools which they would prefer to their actual assignment because they erroneously think they will not be accepted. This again suggests that in real world settings, more top applicants would be selected into top schools if applicants would have better information on their admission probabilities before applying.

\citet{wu2014} use empirical variation in assignment mechanisms in China to study the effect of the mechanism used on applicant selection at a top business school. As should be expected, selection on the actual application score is attenuated when when applicants have an incentive to apply only to programs that would accept them, but are uncertain about which programs these would be. Interestingly, \citeauthor{wu2014} find that this mechanism nevertheless leads to stronger selection on applicants' subsequent university grades, likely because the application score is a bad predictor of subsequent grades, and because the mechanism forces applicants to make strategic decisions based on their more accurate private assessments of their abilities instead.

In the present paper, I use Finnish higher education clearinghouse data to evaluate the role of the Finnish polytechnic assignment mechanism in selecting applicants. Though carried out in a similar spirit, a notable difference with the work of \citeauthor{carvalho2018}\ and \citeauthor{wu2014} is that I study the application system in its entirety rather than analyzing the assignment to one or two programs only. By using the panel structure of the data, I show that the Finnish polytechnic assignment effectively attenuates selection by unnecessarily rejecting top applicants. 

The selection effects of educational assignment should be placed in a wider literature on the institutional determinants of educational selection, in particular with respect to applications and admissions. \citet{hoxby2012}, \citet{hoxby2015}, \citet{dillon2017} and \citet{dynarski2018} for example show that many low income, high ability students unnecessarily fail to apply to selective colleges in the US, while \citet{cohn2004}, \citet{machin2005} and \citet{cliffordson2006} show how the choice of selection criteria affects the selection of applicants into types and levels of education. The assignment mechanism affects both the application and admission phases of the assignment process, and it should not be treated separately from other determinants of educational selection.

The Finnish polytechnic assignment mechanism not only fails to select, but it arguably also fails to assign. When top applicants are rejected, they will re-apply in later years, and even assigned applicants have an incentive to apply to a more preferred program in a later year. As a consequence, queues into higher education are long, and the graduation age is high. Improvements to the mechanism may thus simultaneously solve multiple problems in Finnish higher education that have thus far mainly been viewed through the lens of public economics. 

\section{Finnish higher education admissions}

Finnish higher education is provided by polytechnics and universities, with students typically graduating with a bachelor's degree from the former, and with a master's from the latter.  Higher education is free of charge for the large majority of students, and student benefits are relatively generous. There are substantial lifetime income differences between individuals with different levels of education, also after taxes and transfers \citep{koerselman2014}.

All higher education applications are made to a national clearinghouse. Polytechnic admission decisions are generally made centrally by the clearinghouse itself, while university admission decisions are generally not. In this paper, I analyze the 2011 centralized assignment of polytechnic applicants, in which 50894 polytechnic applicants applied to 16655 seats in 440 programs, divided over 8 fields.

Applicants are admitted according to a program-specific admission score which is mainly based on applicants' matriculation exam grade point averages as well as on their entrance exam results. The weights assigned to different matriculation exam subjects are typically shared within each field, and entrance exams tend to be shared as well. Extra points are awarded for the first listed choice, as well as for factors like relevant labor market experience. The relative weight of the different admission score components in determining the admission score can be seen from Table \ref{decomposition}.

\begin{table}[h]
	\caption{The relative importance of different admission score components.\label{decomposition}}
\begin{footnotesize}
\begin{tabularx}{\columnwidth}{X r}
					& Effective weight\\
\hline
Matriculation exam GPA     &  0.28\\
Entrance exam             &  0.43\\
Program listed first   &  0.05\\
Residual variance      &  0.23\\
\hline
\multicolumn{2}{p{0.97\columnwidth}}{Notes: The table shows the standardized square roots of the variance components of the admission score. Note that a comparison of component means would be uninformative. For example, a score component with positive mean, but with zero variance across applicants would not affect the priority ordering.}\\
\end{tabularx}
\end{footnotesize}
\end{table}
 
The application process starts in March, when applicants can rank up to four programs in order of preference. Entrance exams are mostly held in May or June. Though applicants receive good indications of their matriculation exam grades before they apply, and though they may be aware of previous years' admission score cut-offs, they necessarily learn their entrance exam scores only after choosing where to apply and which entrance exams to take, adding a considerable degree of uncertainty to their application.

After the entrance exams have been graded, applicants are assigned to programs through a centrally run program-proposing deferred acceptance algorithm, each applicant either being admitted to a single program, or not being assigned at all. Assigned applicants then either accept their seat, or reject it. A much smaller second round of offers is sent out by the programs themselves to make up for first-round rejections. The second round of the process ends at the start of the fall term in September.

The use of program-proposing deferred acceptance is remarkable since conditional on submitted applications and realized admission scores it creates the program optimal stable assignment, i.e.\ the stable assignment which maximizes the quality of admitted applicants rather than applicants' welfare. Given the primacy of applicant over program welfare in the literature, a natural first question to ask is thus whether there is more than one stable assignment conditional on applications and scores, and whether there therefore is a different stable assignment which would redistribute welfare from programs to applicants. If the differences between the two assignments are large, the question which assignment to choose may be of political relevance.

Only about a third of polytechnic applicants are admitted each year. It is however not the third of applicants with the highest grades that is admitted, or even the third of applicants with the highest admission scores. Table \ref{applicationefficiency} shows the proportion of applicants ranking in the top third, middle third, and bottom third of the programs they applied to who remained completely unassigned at the end of the main application round of 2011. When classifying applicants three groups based on their program-specific matriculation exam GPA, as many as 54\% of top third applicants remain unassigned anywhere. Even using the actual admission score, 34\% of top third applicants remain unassigned.

\begin{table}[h]
	\caption{Proportion of applicants not assigned to any program by mean priority rank. \label{applicationefficiency}}
\begin{footnotesize}
\begin{tabularx}{\columnwidth}{X r r r}
	& \multicolumn{3}{c}{applicant priority tercile group}\\
Priority criterion	& highest third & middle third & lowest third\\
	\hline
	Matriculation exam&0.54&0.69&0.80\\ 
Admission score&0.34&0.76&0.95\\ 
 [2pt]
	\hline
\multicolumn{4}{p{0.97\columnwidth}}{Notes: The table shows the proportion of applicants who were rejected by every single program they applied to. Applicants are classified into three groups by their average rank across the programs they applied to. Even among applicants who on average rank among the top third of applicants at the programs they apply to, large numbers remain completely unassigned.}\\
\end{tabularx}
\end{footnotesize}
\end{table}

Though it does not follow strictly, the low selectiveness on the admission score is a strong indication that there are large numbers of unassigned applicants which programs would admit over their actual assignment. One reason that top applicants are not assigned to these programs is that they do not apply to other programs than the ones they are rejected from. Table \ref{applicationstatistics} shows descriptive statistics on the application level, i.e.\ for unique combinations of applicants and programs. In 50894 applications, a program was listed first, in 40532 a program was listed second, in 30443 third, and in 19048 fourth. Only 19048 applicants thus applied to the maximum number of four programs. The average number of programs applied to is 2.77.

\begin{table}[t]
	\caption{Statistics on individual program applications.\label{applicationstatistics}}
\begin{footnotesize}
\begin{tabularx}{\columnwidth}{X r r r r}
	& \multicolumn{4}{c}{program rank in submitted list}\\
	& listed $1^{st}$& listed $2^{nd}$& listed $3^{rd}$& listed $4^{th}$\\
	\hline
				 Number of applications&50894&40532&30443&19048\\ 
 [2pt]
				 Proportion with entrance exam taken&0.59&0.48&0.45&0.43\\ 
Proportion admitted&0.27&0.04&0.03&0.03\\ 
 [2pt]
\hline
\end{tabularx}
\end{footnotesize}
\end{table}

Even if applicants do apply to more than one program, their admission chances are relatively low for programs listed second, third and fourth, with the probability of being assigned to a program being 27\% for the program listed first, but only between three and four per cent for programs listed lower. This is partly due to the extra points given for the first listed program, but is probably also related to applicants' strategic choices on which entrance exams to take. The second row of Table \ref{applicationstatistics} shows that in 59\% of applications in which the applicant listed the program first, the applicant had also taken a valid entrance exam for that program. This proportion is lower for programs listed lower, and taken together the figures suggest both that participating in an entrance exam is costly to the applicant, and that applicants concentrate their efforts on the exam of their first listed program. Furthermore, applicants may refrain from applying to additional programs because they may not able to gain admission to those programs without additional effort that would be better spent on their first listed application.

It may be unavoidable that some top applicants are rejected from some programs, but of importance is why they are not then assigned to a different program instead. One possibility is that applicants do not apply to any other programs because they find all other programs unacceptable. In that case, the rejection of top applicants is an almost necessary consequence of applicants' preferences, and perhaps not easily changed by policy. Another possibility is that top applicants do find other programs acceptable, but do not apply to them, for example because they deem their admission probabilities low for programs not listed first. A second question I try to answer is whether poor selectivity is a necessary consequence of applicants preferences, or whether there are unassigned top applicants that can be shown to find other programs than the ones applied to acceptable. 

Some top applicants who are rejected from their first listed choice do apply to more than one program, but are also rejected from the other programs they applied to. This may be the case because they only applied to very selective programs, but for example also because they are disadvantaged in applications to programs they did not list first. The extra points awarded for the first listed choice effectively force programs to reject otherwise higher scoring applicants in favor of lower scoring applicants who listed the program first. The use of entrance exams additionally forces programs to prioritize applicants with a poor exam result over those who did not attempt the exam. Both factors are likely to reduce selection on the skills which the matriculation and entrance exams have been designed to test. A third question is whether rejected top applicants who do apply to more than one program are rejected from other programs solely because they did not receive the extra points for the first choice, and whether they were rejected because they did not take the entrance exam.

\section{Data and methods}
I use data covering all applications to Finnish polytechnics that were made through the centralized clearing mechanism during the fall application rounds of the years 2011, 2012, and 2013. The data also contain information on university applications, even if admission decisions were typically made in a decentralized fashion for university programs. For polytechnics, the data contain the programs applied to in stated order of preference, matriculation exam grades, entrance exam scores, and the composite application score. The final state of the assignment algorithm is also available, indicating which applicants were assigned seats where, and which assigned seats were accepted by applicants. I evaluate the 2011 assignment in my analysis, using information from re-applications in 2012 and 2013 to create counterfactual assignments for 2011. I limit the analysis to the main group of polytechnic applicants who applied on the basis of a general high school degree, and thus exclude applicants who entered after completing a vocational program.

Regrettably, information on the program identifiers and quota used by the algorithm is not available. I therefore use combinations of polytechnic name and program name as proxies of program identifiers and the number of simultaneous offers made by each program as a proxy for program quota. I can replicate about 98\% of application decisions by applying a standard program-proposing deferred acceptance algorithm to the admission scores. The slight discrepancy between actual and replicated application decisions may be explained not only by inexact program identifiers, but also by applicants' failure to fulfill secondary entrance criteria not captured in the admission score, for example when failing to prove their competency in the language of instruction at the polytechnic applied to. I use the replicated assignment as the benchmark to compare counterfactual assignments to, but base all descriptive statistics on the actual assignment.

The first question I answer is whether there exists more than one stable and therefore legal assignment conditional on submitted applications and admission scores. To do so, I apply an applicant-proposing deferred acceptance algorithm to the empirical applications and admission scores, and compare the resulting applicant optimal stable assignment to the program optimal stable assignment. Since the program and applicant optimal stable assignments form extremes of the set of stable assignments, when the two coincide, there is only one stable assignment consistent with submitted applications and realized admission scores. 

Second, even in terms of the actual assignment criteria, many top applicants remain unassigned each year. Are there programs which these applicants would have found acceptable, and would have been admitted to if it was not for the fact that they did not apply? Are there applicants who did apply to additional programs, but who were rejected because of the extra points awarded for their first choice, or because they did not take the entrance exam?

I observe applicants not only in the 2011 application round that I analyze, but also in the application rounds of 2012 and 2013, when many 2011 applicants re-enter the application system. Under the assumption that programs acceptable to an applicant in 2012 and 2013 would also have been acceptable to the applicant in 2011 conditional on being rejected from the programs applied to in 2011, I use the panel structure of the data to create counterfactual applications in 2011, adding to the end of the submitted applications in 2011 the same applicants' 2012 applications to programs not applied to in 2011, and to the end of the resulting list 2013 applications to programs not applied to in 2011 or 2012. I furthermore create two counterfactual sets of admission scores. In the first, I mechanically remove the extra points awarded for the first listed choice. In the second, I additionally make the first entrance exam taken within each field valid for all applications within that field in which the applicant does not have a valid entrance exam result. An applicant who for example applied to an engineering program in 2011, 2012 and 2013, but took an engineering entrance exam only in 2012 and 2013, would thus have the 2012 exam result added to her 2011 engineering admission score.

Using all combinations of empirical and counterfactual applications and empirical and counterfactual admission scores, I re-apply the deferred acceptance algorithm to create five counterfactual assignments which I then compare to the actual assignment in terms of the number of differently assigned applicants. If the number of reassigned applicants is large, the design of the admissions system can be thought to have a large effect on the assignment. To determine whether the counterfactual assignments are associated with changes in applicant quality, I evaluate quality by a measure which is not directly affected by the counterfactual changes to admission scores themselves, and which is known for all combinations of applicant and program: the field-specific matriculation GPA rank among the population of applicants. 

It is possible that applicants' incentives to concentrate their efforts on a (very) low number of applications helps them signal their motivation to complete the specific program(s) they apply to. I estimate a series of linear probability models using assigned applicants' submitted program ranking and assigned applicants' participation in a relevant entrance exam to predict whether they will accept the assigned seat, and whether they will re-enter the application system to apply for a different program in 2012 or 2013. Since both predictors and outcomes may differ systematically between applicants with different skill levels, as well as between differentially selective programs, I control for an application-specific adjusted admission score from which I have subtracted the points awarded for the first listed program and the entrance exam result. I also control for the program-specific admission score of its lowest scoring admitted applicant. Because admission criteria differ between fields of study, I furthermore include as controls all interactions of these two variables with field of study.

\section{Results}

I start by applying an applicant-proposing rather than a program-proposing deferred acceptance algorithm to the empirical applications and priority orderings. I find that not a single applicant is assigned to a different program under the the applicant-proposing assignment than under the program-proposing assignment. Conditional on submitted applications, there is thus only a single way to assign applicants which is stable conditional on actual applications and admission scores, i.e.\ which does not violate applicants' priority rights for the programs they applied to. This finding is in line with the earlier literature, and simulations not shown here suggest that large differences between the two extreme assignments are extremely unlikely to occur in higher education admissions.\footnote{The size of the set of stable assignments can be large under some idealized conditions \citep{pittel1989}, and small under others \citep{azevedo2016}. Typically however, the set is small \citep[e.g.][]{roth1999}. There are various known circumstances which tend to produce this result: when the number of applicants differs from the number of seats \citep{ashlagi2017}, when applicants each apply to a limited number of programs only \citep{roth1999, immorlica2005,kojima2009}, and when either program or applicant preferences are correlated with each other \citep{roth1999, holzman2014, ashlagi2017}. These circumstances are all common in higher education admissions.}

Applicants only apply to a small number of programs each, and many top applicants remain unassigned anywhere. This could be a necessary consequence of programs other than the ones applied to being unacceptable to applicants. Many rejected applicants however re-apply in later years, and not always to the same programs. This suggests that these other programs are also acceptable to the applicant, at least conditional on being rejected from the programs originally applied to. Table \ref{improvements} shows how the assignment changes if we were to process 2012 and 2013 applications to these other programs already in 2011. As can be seen from the first two rows of the table, this increases the average number of programs applied to from  
2.77 
to
3.72,
and causes
10\%
of applicants to receive a different assignment outcome. As a consequence, the average field-specific matriculation GPA of assigned applicants improves by  
1.45 
percentile ranks. Since only about a third of applicants is assigned anywhere, a different assignment outcome for 10\% of applicants is not insubstantial. 

\begin{figure}[t]
\begin{center}
\caption{The grade rank distribution of assigned applicants. \label{fig:qualityfigure}}
\includegraphics[width=0.9\columnwidth]{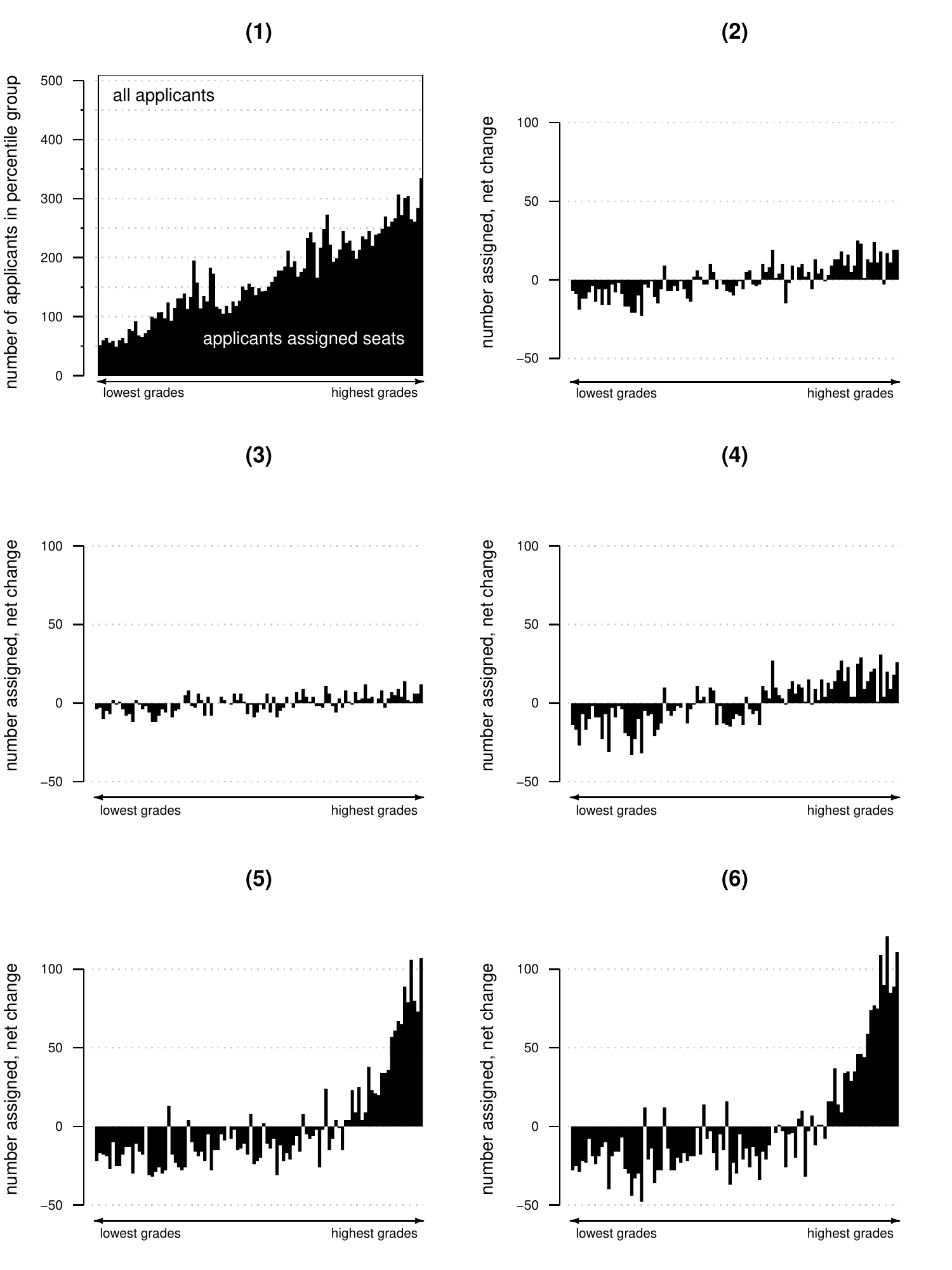}
\end{center}
	\footnotesize{Panel (1) shows the field-specific matriculation GPA rank distribution of assigned applicants. Because applicants' grades are weighted somewhat differently in different fields, there is not a single correct way to rank unassigned applicants, but a uniform rank distribution has been added to the figure for illustrative purposes. Panels (2) through (6) show the net change in the number of assigned applicants by grade percentile group for the five counterfactual assignments of Table \ref{improvements}.}
\end{figure}

The actual GPA rank distribution of assigned applicants is shown in the first panel of Figure \ref{fig:qualityfigure}, in which assigned applicants have been sorted into 100 bins by their field-specific GPA rank at the program they were assigned to. As can be seen from the figure, even among applicants who are in the top 1\% of the GPA distribution, about one third remains unassigned.\footnote{Since applicants rank differently at different programs, there is not a single correct way to rank unassigned applicants. For illustrative purposes, an uniform distribution of GPA rank is used as the implicit denominator. Alternate ways of ranking unassigned applicants produce very similar denominators.} The second panel shows the net change in assigned applicants in each percentile group when also including the applicants' future applications. The change is relatively linear, with fewer low scoring applicants being admitted, small net changes in the middle of the distribution, and more high scoring applicants being admitted.

\begin{table}[h]
	\caption{Counterfactual improvements to the matriculation grades of admitted applicants.\label{improvements}}
\begin{footnotesize}
\begin{tabularx}{\columnwidth}{X r r r}
	\hline
& applications  & differently & total\\
& per           & assigned    & pct.\ rank\\
& applicant     & applicants & imprvmnt\\
\hline
\emph{Original admission score}&&&\\
(1) Original applications&
2.77 &
&
\\
(2) Extended applications$^a$&
3.72 &
10\%&
+1.45\\[4pt]

\emph{First choice points removed}&&&\\
(3) Original applications&
2.77 &
5\%&
+0.56\\
(4) Extended applications$^a$&
3.72 &
15\%&
+1.87\\[4pt]

\emph{First exam furthermore universally valid}$^{b}$&&&\\
(5) Original applications&
2.77 &
14\%&
+3.97\\
(6) Extended applications$^a$&
3.72 &
24\%&
+4.59\\[4pt]
\hline
	\multicolumn{4}{p{0.97\columnwidth}}{Notes: The table shows the changes to the assignment resulting from counterfactual changes to submitted applications and to the priority ordering. Percentages and percentile ranks are calculated across all applicants. The empirical mean percentile rank of admitted applicants is 61.63, scaled so that a higher percentile rank implies a higher GPA. $a$: Later year applications added to bottom of earlier year rank ordered list. $b$: The first entrance exam taken in each field considered valid for all applications in that field for which no exam was taken.}\\
\end{tabularx}
\end{footnotesize}
\end{table}

Table Rows (3) and (5) show the assignment differences from the original assignment when the extra points awarded to the program listed first are removed, and when entrance exams results are furthermore extended across years respectively. Removing the extra points allows programs to reject applicants with an otherwise lower score who listed the program first in favor of applicants with a higher score who did not. 5\% of applicants are assigned differently when the extra points are removed, leading to a 0.56 mean percentile rank improvement in the field-specific matriculation GPA of admitted applicants. When additionally adding their first entrance exam score to applicants' admission scores for applications within the same field in which they did not take an entrance exam, 14\% of applicants are reassigned compared to the original assignment, causing a 3.97 percentile rank improvement in the grades of accepted applicants. The distributional changes can be seen in Panels (3) and (5) of Figure \ref{fig:qualityfigure} respectively. Notably, the inclusion of future entrance exam scores greatly increases the number of assigned applicants from the top fifth, and reduces the number of assigned applicants from the lower four fifths of the GPA rank distribution. 

Table Row (4) shows the effects of simultaneously extending application lists with programs not originally applied to and removing the extra points for the first application. Changes to the assignment are larger than when making either change in isolation. Row (6) shows the effects of additionally making entrance exams valid across years. When doing that, as many as 24\% of applicants receive a different assignment outcome. This is a large change for a system that only admits a third of its applicants. The associated difference in the mean percentile rank of assigned applicants is 4.59. If the change would have come about by rejecting 12 percentage points of previously assigned applicants, and admitting 12 percentage points of previously unassigned applicants, each newly assigned applicant would have to have an average 13 percentile ranks higher grades than the applicant s/he replaced. The actual distributional changes can be seen from the corresponding Panels (4) and (6) in Figure \ref{fig:qualityfigure}, where we again see that the large improvements in the grades of assigned applicants when including future entrance exam results come about through large positive net changes in the top one fifth of the distribution, and negative net changes in the rest. There thus exists a large quantity of top fifth applicants who would have been admitted had they immediately applied to the programs they were going to apply to in the future, had they not been disadvantaged by the extra points other applicants received for their first listed choice, and had they obtained their future entrance exam result immediately.

\afterpage\clearpage

The points implicitly and explicitly awarded for applicants' strategic choices come at a cost in terms of selection on matriculation and entrance exam performance, but applicants' strategic choices may be associated with a valuable trait: that the applicant is serious about the application. The first three columns of Table \ref{predictor} contain estimates from a series of linear probability models, regressing whether the admitted applicant accepted the assigned seat on the program's position in the submitted application list as well as on whether the applicant participated in the entrance exam. Column (1) shows that conditional on each other, the applicant is 11.8 percentage points less likely to accept the assigned seat when listing a program second instead of first, 11.2 percentage points less likely when listing it third, and 18.8 percentage points less likely when listing it fourth, and that the applicant is 25 percentage points less likely to accept the assigned seat when s/he did not take a relevant entrance exam. When controlling for the applicant's score and the selectivity of the program applied to in Column (2), and for both variables interacted with field of study in Column (3), the entrance exam coefficient is reduced by a fifth, but the estimates are otherwise substantially unchanged. All these estimates are relatively precise, and significantly different from zero.

Columns (4) through (6) show estimates from a similar set of models, but with the outcome now being whether the applicant re-entered the higher education application system in 2012 or 2013. We see that applicants who were admitted at a program they did not list first were 10 to 20 percentage points more likely to re-enter the application system in a future year. Applicants who had taken an entrance exam for the program they were accepted to were less likely to re-apply in a future year, but although the effect is of the expected sign, the estimate clears the 5\% significance threshold only in Column (6). Since all other estimates are large and statistically distinguishable from zero however, considered together, the elements of the admission score which are dependent on applicants' strategic choices are clearly jointly predictive of applicants' subsequent decisions on whether to accept the assigned seat, and whether to try to gain entry to another program in a future year.

\afterpage{
	\clearpage
	\begin{landscape}
	\begin{table}
	\centering
		\caption{Listed preference order and entrance exam participation among applicants admitted in 2011 as predictors of their subsequent decisions. \label{predictor}}
	\begin{footnotesize}
	\begin{tabular}{l r r r r r r}
		& (1) & (2) & (3) & (4) & (5) & (6)\\
	Dependent variable			& \multicolumn{3}{c}{Accepted seat} & \multicolumn{3}{c}{Re-applied later} \\
	\hline  
		Listed program second     	& -0.118 
						& -0.122 
						& -0.124 
						& 0.111 
						& 0.113 
						& 0.109\\
						   
						& (0.010) 
						& (0.010) 
						& (0.010) 
						& (0.012) 
						& (0.012) 
						& (0.012)\\[2pt]
						
		Listed program third     	& -0.112 
						& -0.115 
						& -0.122 
						& 0.190 
						& 0.191 
						& 0.187\\
		 
						& (0.013) 
						& (0.013) 
						& (0.013) 
						& (0.016) 
						& (0.016) 
						& (0.016)\\[2pt]

		Listed program fourth     	& -0.188 
						& -0.194 
						& -0.204 
						& 0.215 
						& 0.216 
						& 0.210\\
			 
						& (0.018) 
						& (0.018) 
						& (0.018) 
						& (0.021) 
						& (0.021) 
						& (0.021)\\[2pt]
				
		Took entrance exam	     	& 0.250 
						& 0.207 
						& 0.202 
						& -0.053 
						& -0.027 
						& -0.124\\
	 
						& (0.037) 
						& (0.037) 
						& (0.052) 
						& (0.044) 
						& (0.044) 
						& (0.062)\\[4pt]

	Adjusted applicant score 		& \textsc{  }          & \textsc{yes}       & \textsc{yes}        & \textsc{  } & \textsc{yes}        & \textsc{yes} \\              
	Program acceptance threshold  		& \textsc{  }          & \textsc{yes}       & \textsc{yes}        & \textsc{  } & \textsc{yes}        & \textsc{yes} \\              
	Field of study				& \textsc{  }          & \textsc{  }        & \textsc{yes}        & \textsc{  } & \textsc{  }        & \textsc{yes} \\              
	Adjusted applicant score $\times$ field of study	& \textsc{  }          & \textsc{  }        & \textsc{yes}        & \textsc{  } & \textsc{  }        & \textsc{yes} \\              
	Program acceptance threshold $\times$ field of study & \textsc{  }     & \textsc{  }        & \textsc{yes}        & \textsc{  } & \textsc{  }        & \textsc{yes} \\              
	\hline
	Mean dependent variable	& 0.813 & 0.813 & 0.813 & 0.304 & 0.304 & 0.304\\
	$n$			& 16655 & 16655 & 16655 & 16655 & 16655 & 16655\\
	\hline  
		\multicolumn{7}{p{16.2cm}}{Notes: The table shows estimates from a linear probability model on admitted applicants. The adjusted applicant score is the admission score with the entrance exam result and the extra points for the first listed choice subtracted. Standard errors have been added in parentheses.} \\
	\end{tabular}
	\end{footnotesize}
	\end{table}
	\end{landscape}
	\clearpage
}

\section{Discussion}
I use a panel of Finnish higher education clearinghouse data to evaluate the selection of applicants into polytechnics. Many top applicants remain unassigned by the mechanism, both when top applicants are defined in terms of their matriculation exam GPAs, and when defined in terms of the actual admission score that determines admissions.

The number of programs each applicant applies to is low, and when a top applicant is rejected from a competitive program, there may not be another program which the applicant can be assigned to. While some applicants may find all programs not applied to unacceptable, large numbers of applicants can be observed to apply to at least one different program in one of the two subsequent years. If the applicants would have applied to those programs immediately, programs would have a wider choice of applicants, and selection on observable skills would improve considerably.

The application system limits the number of programs applicants can effectively apply to. There is a hard limit of four programs, but extra application points are awarded for the first listed choice, and applicants furthermore have to make a strategic decision on which entrance exams to take. As a consequence, few applicants are admitted to a program that they have not listed first. This not only discourages applicants from applying to multiple programs, but also enforces selection on strategic choices rather than on skills. Many top applicants do apply to more than one program, but are rejected because they do not list that program first, or because they have not taken the relevant entrance exam for that program. There exists a large quantity of top fifth applicants who would have been admitted had they immediately applied to the programs they were going to apply to in the future, had they not been disadvantaged by the extra points other applicants received for their first listed choice, and had they obtained their future entrance exam result immediately.

The Boston school choice mechanism can create an all-or-nothing dynamic in submitted applications. When combined with uncertainty on the set of programs that would be willing to admit any specific applicant, this can effectively create a price for competitive programs in terms of the risk applicants are willing to take to not be assigned to it \citep[cf.][]{abdulkadirouglu2011}. As a consequence, in equilibrium applicants apply to a more competitive program only if they value attendance at that program unusually highly. The result can be seen as a kind of selection on otherwise unobservable characteristics, and it should be expected to be present in the Finnish polytechnic assignment as well. Since selecting the most motivated applicants into each program can benefit both the applicants themselves and the programs they attend, this is an argument in favor of the mechanism.

An important way in which the Finnish polytechnic assignment differs from the Boston school assignment is however that applicants who unsuccessfully apply to a competitive program are not immediately and permanently assigned to a different program. Rather, they can enter the application system multiple years in a row. Though the Finnish polytechnic assignment creates a stable assignment with respect to a single year's submitted preferences and realized admission scores, the assignment is evidently not stable with respect to future submitted applications and future admission scores. Applicants can receive different assignments in different years, and thus generally have an incentive to re-apply. 

Re-applications inflate the yearly applicant numbers, and effectively create queues into higher education. Any Boston-like benefits from applicants being assigned to programs they value unusually highly are in the Finnish case likely to be more than outweighed by the number of years they have to re-apply before they are assigned to it. Finnish students have one of the highest graduation ages in the OECD \citep[][Chart A3.1]{OECD2014}, and although various attempts have been made to encourage a more timely graduation \citep[cf. e.g.][]{hamalainen2017}, at least part of the reason for Finnish students' high graduation age is the high age at which they enroll in the program they will graduate from.

Even admitted applicants frequently refrain from accepting their seat, or re-apply to a different program in a later year. Since the admission score components which depend on applicants' strategic choices are predictive of these outcomes, it may seem attractive to argue that retention problems would be worse if strategic choices would not be used as selection criteria. The observed relationship between strategic choices and subsequent behavior is however not informative of applicants' behavior under different application rules. On the contrary, it seems likely that the poor performance of the assignment mechanism is a main reason why so many applicants refrain from attending and completing the program they were assigned to in the first place.

Rather than trying to predict which applicants will re-apply, and rather than trying to stop applicants from re-applying altogether, it should be possible to immediately assign most applicants to their most preferred feasible program. One way to try to achieve this would be to create a true applicant optimal stable mechanism by removing selection on strategic choices, allowing and encouraging applicants to list a large number of programs in their true order of preference. This may however not work as well as theory would predict. Applying to many programs at once may be psychologically costly, and applicants may erroneously discount the possibility that they would be admitted at a competitive program. Applicants could therefore alternatively be given reliable information on the set of programs that would be willing to admit them already before they make their final decision on where to apply, for example by placing the entrance exam earlier in the application process \citep[cf.][]{fack2018,carvalho2018,chen2019}. 

It should be noted that in the Finnish polytechnic assignment, entrance exams can be retaken in subsequent years, and applicants thus typically have an incentive to re-apply even when assigned to their originally most preferred feasible program. Though fairness concerns may require that any exam can be retaken, selection criteria should be constructed in a way that avoids unnecessary year-to-year variation in application scores if the assignment is to be long-term stable. 

\section*{Acknowledgments}
I am grateful to Angela Djupsjöbacka, Hannu Vartiainen, Roope Uusitalo, Matthias Strifler, Mikko Salonen, Hannu Karhunen, Jouni Helin, Aleksi Kalenius, Antti Kauhanen, as well as to participants at various workshops and conferences for their kind help and advice. Part of the work was carried out at Åbo Akademi University, at the University of Helsinki, at Stockholm University, and at the University of Zaragoza.

\section*{Funding sources}
This work was supported by the Finnish Institute for Educational Research [Cygnaeus Scholarly Fellowship 2.0, 2019]. 

\section*{Declarations of interest}
None.

\bibliographystyle{model5-names}
\bibliography{polytechnics}

\begin{thebibliography}{33}
\expandafter\ifx\csname natexlab\endcsname\relax\def\natexlab#1{#1}\fi
\providecommand{\bibinfo}[2]{#2}
\ifx\xfnm\relax \def\xfnm[#1]{\unskip,\space#1}\fi
%Type = Article
\bibitem[{Abdulkadiro{\u{g}}lu et~al.(2011)Abdulkadiro{\u{g}}lu, Che \&
  Yasuda}]{abdulkadirouglu2011}
\bibinfo{author}{Abdulkadiro{\u{g}}lu, A.}, \bibinfo{author}{Che, Y.-K.}, \&
  \bibinfo{author}{Yasuda, Y.} (\bibinfo{year}{2011}).
\newblock Resolving Conflicting Preferences in School Choice: The ``Boston
  Mechanism'' Reconsidered.
\newblock {\it \bibinfo{journal}{American Economic Review}\/},  {\it
  \bibinfo{volume}{101}\/}, \bibinfo{pages}{399--410}.
%Type = Article
\bibitem[{Abdulkadiro{\u{g}}lu et~al.(2005{\natexlab{a}})Abdulkadiro{\u{g}}lu,
  Pathak \& Roth}]{abdulkadiroglu2005nyc}
\bibinfo{author}{Abdulkadiro{\u{g}}lu, A.}, \bibinfo{author}{Pathak, P.~A.}, \&
  \bibinfo{author}{Roth, A.~E.} (\bibinfo{year}{2005}{\natexlab{a}}).
\newblock The New York City High School Match.
\newblock {\it \bibinfo{journal}{American Economic Review}\/},  {\it
  \bibinfo{volume}{95}\/}, \bibinfo{pages}{364--367}.
%Type = Article
\bibitem[{Abdulkadiro{\u{g}}lu et~al.(2009)Abdulkadiro{\u{g}}lu, Pathak \&
  Roth}]{abdulkadiroglu2009}
\bibinfo{author}{Abdulkadiro{\u{g}}lu, A.}, \bibinfo{author}{Pathak, P.~A.}, \&
  \bibinfo{author}{Roth, A.~E.} (\bibinfo{year}{2009}).
\newblock Strategy-proofness versus Efficiency in Matching with Indifferences:
  Redesigning the NYC High School Match.
\newblock {\it \bibinfo{journal}{American Economic Review}\/},  {\it
  \bibinfo{volume}{99}\/}, \bibinfo{pages}{1954--78}.
%Type = Article
\bibitem[{Abdulkadiro{\u{g}}lu et~al.(2005{\natexlab{b}})Abdulkadiro{\u{g}}lu,
  Pathak, Roth \& S{\"o}nmez}]{abdulkadiroglu2005boston}
\bibinfo{author}{Abdulkadiro{\u{g}}lu, A.}, \bibinfo{author}{Pathak, P.~A.},
  \bibinfo{author}{Roth, A.~E.}, \& \bibinfo{author}{S{\"o}nmez, T.}
  (\bibinfo{year}{2005}{\natexlab{b}}).
\newblock The Boston Public School Match.
\newblock {\it \bibinfo{journal}{American Economic Review}\/},  {\it
  \bibinfo{volume}{95}\/}, \bibinfo{pages}{368--371}.
%Type = Article
\bibitem[{Abdulkadiroğlu \& S\"{o}nmez(2003)}]{AbdulkadirogluSonmez2003}
\bibinfo{author}{Abdulkadiroğlu, A.}, \& \bibinfo{author}{S\"{o}nmez, T.}
  (\bibinfo{year}{2003}).
\newblock {School Choice: A Mechanism Design Approach}.
\newblock {\it \bibinfo{journal}{American Economic Review}\/},  {\it
  \bibinfo{volume}{93}\/}, \bibinfo{pages}{729--747}.
%Type = Article
\bibitem[{Ashlagi et~al.(2017)Ashlagi, Kanoria \& Leshno}]{ashlagi2017}
\bibinfo{author}{Ashlagi, I.}, \bibinfo{author}{Kanoria, Y.}, \&
  \bibinfo{author}{Leshno, J.~D.} (\bibinfo{year}{2017}).
\newblock Unbalanced Random Matching Markets: The Stark Effect of Competition.
\newblock {\it \bibinfo{journal}{Journal of Political Economy}\/},  {\it
  \bibinfo{volume}{125}\/}, \bibinfo{pages}{69--98}.
%Type = Article
\bibitem[{Azevedo \& Leshno(2016)}]{azevedo2016}
\bibinfo{author}{Azevedo, E.~M.}, \& \bibinfo{author}{Leshno, J.~D.}
  (\bibinfo{year}{2016}).
\newblock A Supply and Demand Framework for Two-Sided Matching Markets.
\newblock {\it \bibinfo{journal}{Journal of Political Economy}\/},  {\it
  \bibinfo{volume}{124}\/}, \bibinfo{pages}{1235--1268}.
%Type = Article
\bibitem[{Balinski \& S{\"o}nmez(1999)}]{balinski1999}
\bibinfo{author}{Balinski, M.}, \& \bibinfo{author}{S{\"o}nmez, T.}
  (\bibinfo{year}{1999}).
\newblock A Tale of Two Mechanisms: Student Placement.
\newblock {\it \bibinfo{journal}{Journal of Economic theory}\/},  {\it
  \bibinfo{volume}{84}\/}, \bibinfo{pages}{73--94}.
%Type = Misc
\bibitem[{Carvalho et~al.(2018)Carvalho, Magnac \& Xiong}]{carvalho2018}
\bibinfo{author}{Carvalho, J.-R.}, \bibinfo{author}{Magnac, T.}, \&
  \bibinfo{author}{Xiong, Q.} (\bibinfo{year}{2018}).
\newblock College Choice, Selection and Allocation Mechanisms: A Structural
  Empirical Analysis.
%Type = Article
\bibitem[{Chen \& Pereyra(2019)}]{chen2019}
\bibinfo{author}{Chen, L.}, \& \bibinfo{author}{Pereyra, J.~S.}
  (\bibinfo{year}{2019}).
\newblock Self-selection in School Choice.
\newblock {\it \bibinfo{journal}{Games and Economic Behavior}\/},  {\it
  \bibinfo{volume}{117}\/}, \bibinfo{pages}{59 -- 81}.
%Type = Article
\bibitem[{Cliffordson \& Askling(2006)}]{cliffordson2006}
\bibinfo{author}{Cliffordson, C.}, \& \bibinfo{author}{Askling, B.}
  (\bibinfo{year}{2006}).
\newblock Different Grounds for Admission: Its Effects on Recruitment and
  Achievement in Medical Education.
\newblock {\it \bibinfo{journal}{Scandinavian Journal of Educational
  Research}\/},  {\it \bibinfo{volume}{50}\/}, \bibinfo{pages}{45--62}.
%Type = Article
\bibitem[{Cohn et~al.(2004)Cohn, Cohn, Balch \& Bradley~Jr}]{cohn2004}
\bibinfo{author}{Cohn, E.}, \bibinfo{author}{Cohn, S.}, \bibinfo{author}{Balch,
  D.~C.}, \& \bibinfo{author}{Bradley~Jr, J.} (\bibinfo{year}{2004}).
\newblock Determinants of Undergraduate GPAs: Sat Scores, High-School GPA and
  High-School Rank.
\newblock {\it \bibinfo{journal}{Economics of Education Review}\/},  {\it
  \bibinfo{volume}{23}\/}, \bibinfo{pages}{577--586}.
%Type = Article
\bibitem[{Dillon \& Smith(2017)}]{dillon2017}
\bibinfo{author}{Dillon, E.~W.}, \& \bibinfo{author}{Smith, J.~A.}
  (\bibinfo{year}{2017}).
\newblock Determinants of the Match Between Student Ability and College
  Quality.
\newblock {\it \bibinfo{journal}{Journal of Labor Economics}\/},  {\it
  \bibinfo{volume}{35}\/}, \bibinfo{pages}{45--66}.
%Type = Article
\bibitem[{Dur et~al.(2018)Dur, Kominers, Pathak \& Sönmez}]{dur2018}
\bibinfo{author}{Dur, U.}, \bibinfo{author}{Kominers, S.~D.},
  \bibinfo{author}{Pathak, P.~A.}, \& \bibinfo{author}{Sönmez, T.}
  (\bibinfo{year}{2018}).
\newblock {Reserve Design: Unintended Consequences and the Demise of Boston’s
  Walk Zones}.
\newblock {\it \bibinfo{journal}{Journal of Political Economy}\/},  {\it
  \bibinfo{volume}{126}\/}, \bibinfo{pages}{2457--2479}.
%Type = Unpublished
\bibitem[{Dynarski et~al.(2018)Dynarski, Libassi, Michelmore \&
  Owen}]{dynarski2018}
\bibinfo{author}{Dynarski, S.}, \bibinfo{author}{Libassi, C.},
  \bibinfo{author}{Michelmore, K.}, \& \bibinfo{author}{Owen, S.}
  (\bibinfo{year}{2018}).
\newblock Closing the Gap: The Effect of a Targeted, Tuition-Free Promise on
  College Choices of High-Achieving, Low-Income Students.
\newblock \bibinfo{note}{National Bureau of Economic Research Working Paper
  25349}.
%Type = Article
\bibitem[{Echenique \& Yenmez(2015)}]{echenique2015}
\bibinfo{author}{Echenique, F.}, \& \bibinfo{author}{Yenmez, M.~B.}
  (\bibinfo{year}{2015}).
\newblock How to Control Controlled School Choice.
\newblock {\it \bibinfo{journal}{American Economic Review}\/},  {\it
  \bibinfo{volume}{105}\/}, \bibinfo{pages}{2679--94}.
%Type = Unpublished
\bibitem[{Fack et~al.(2018)Fack, Grenet \& He}]{fack2018}
\bibinfo{author}{Fack, G.}, \bibinfo{author}{Grenet, J.}, \&
  \bibinfo{author}{He, Y.} (\bibinfo{year}{2018}).
\newblock Beyond Truth-Telling: Preference Estimation with Centralized School
  Choice.
\newblock \bibinfo{note}{American Economic Review, forthcoming}.
%Type = Article
\bibitem[{Gale \& Shapley(1962)}]{GaleShapley1962}
\bibinfo{author}{Gale, D.}, \& \bibinfo{author}{Shapley, L.~S.}
  (\bibinfo{year}{1962}).
\newblock {College Admissions and the Stability of Marriage}.
\newblock {\it \bibinfo{journal}{American Mathematical Monthly}\/},  {\it
  \bibinfo{volume}{69}\/}, \bibinfo{pages}{9--15}.
%Type = Techreport
\bibitem[{H{\"a}m{\"a}l{\"a}inen et~al.(2017)H{\"a}m{\"a}l{\"a}inen, Koerselman
  \& Uusitalo}]{hamalainen2017}
\bibinfo{author}{H{\"a}m{\"a}l{\"a}inen, U.}, \bibinfo{author}{Koerselman, K.},
  \& \bibinfo{author}{Uusitalo, R.} (\bibinfo{year}{2017}).
\newblock {\it \bibinfo{title}{Graduation Incentives Through Conditional
  Student Loan Forgiveness}\/}.
\newblock \bibinfo{type}{Technical Report} IZA Discussion Papers.
%Type = Article
\bibitem[{Holzman \& Samet(2014)}]{holzman2014}
\bibinfo{author}{Holzman, R.}, \& \bibinfo{author}{Samet, D.}
  (\bibinfo{year}{2014}).
\newblock Matching of Like Rank and the Size of the Core in the Marriage
  Problem.
\newblock {\it \bibinfo{journal}{Games and Economic Behavior}\/},  {\it
  \bibinfo{volume}{88}\/}, \bibinfo{pages}{277--285}.
%Type = Techreport
\bibitem[{Hoxby \& Avery(2012)}]{hoxby2012}
\bibinfo{author}{Hoxby, C.~M.}, \& \bibinfo{author}{Avery, C.}
  (\bibinfo{year}{2012}).
\newblock {\it \bibinfo{title}{The Missing ``One-Offs'': The Hidden Supply of
  High-Achieving, Low Income Students}\/}.
\newblock \bibinfo{type}{Technical Report} National Bureau of Economic
  Research.
%Type = Article
\bibitem[{Hoxby \& Turner(2015)}]{hoxby2015}
\bibinfo{author}{Hoxby, C.~M.}, \& \bibinfo{author}{Turner, S.}
  (\bibinfo{year}{2015}).
\newblock What High-Achieving Low-Income Students Know About College.
\newblock {\it \bibinfo{journal}{American Economic Review}\/},  {\it
  \bibinfo{volume}{105}\/}, \bibinfo{pages}{514--17}.
%Type = Inproceedings
\bibitem[{Immorlica \& Mahdian(2005)}]{immorlica2005}
\bibinfo{author}{Immorlica, N.}, \& \bibinfo{author}{Mahdian, M.}
  (\bibinfo{year}{2005}).
\newblock Marriage, Honesty, and Stability.
\newblock In {\it \bibinfo{booktitle}{Proceedings of the sixteenth annual
  ACM-SIAM symposium on Discrete algorithms}\/} (pp. \bibinfo{pages}{53--62}).
\newblock \bibinfo{organization}{Society for Industrial and Applied
  Mathematics}.
%Type = Article
\bibitem[{Kesten(2010)}]{kesten2010}
\bibinfo{author}{Kesten, O.} (\bibinfo{year}{2010}).
\newblock School Choice with Consent.
\newblock {\it \bibinfo{journal}{The Quarterly Journal of Economics}\/},  {\it
  \bibinfo{volume}{125}\/}, \bibinfo{pages}{1297--1348}.
%Type = Article
\bibitem[{Koerselman \& Uusitalo(2014)}]{koerselman2014}
\bibinfo{author}{Koerselman, K.}, \& \bibinfo{author}{Uusitalo, R.}
  (\bibinfo{year}{2014}).
\newblock The Risk and Return of Human Capital Investments.
\newblock {\it \bibinfo{journal}{Labour Economics}\/},  {\it
  \bibinfo{volume}{30}\/}, \bibinfo{pages}{154--163}.
%Type = Article
\bibitem[{Kojima \& Pathak(2009)}]{kojima2009}
\bibinfo{author}{Kojima, F.}, \& \bibinfo{author}{Pathak, P.~A.}
  (\bibinfo{year}{2009}).
\newblock Incentives and Stability in Large Two-Sided Matching Markets.
\newblock {\it \bibinfo{journal}{American Economic Review}\/},  {\it
  \bibinfo{volume}{99}\/}, \bibinfo{pages}{608--27}.
%Type = Article
\bibitem[{Machin \& McNally(2005)}]{machin2005}
\bibinfo{author}{Machin, S.}, \& \bibinfo{author}{McNally, S.}
  (\bibinfo{year}{2005}).
\newblock Gender and Student Achievement in English Schools.
\newblock {\it \bibinfo{journal}{Oxford Review of Economic Policy}\/},  {\it
  \bibinfo{volume}{21}\/}, \bibinfo{pages}{357--372}.
%Type = Techreport
\bibitem[{OECD(2014)}]{OECD2014}
\bibinfo{author}{OECD} (\bibinfo{year}{2014}).
\newblock {\it \bibinfo{title}{Education at a Glance 2014: OECD Indicators}\/}.
\newblock \bibinfo{type}{Technical Report}.
%Type = Inbook
\bibitem[{Pathak(2017)}]{pathak2017}
\bibinfo{author}{Pathak, P.~A.} (\bibinfo{year}{2017}).
\newblock What Really Matters in Designing School Choice Mechanisms.
\newblock In \bibinfo{editor}{B.~Honoré}, \bibinfo{editor}{A.~Pakes},
  \bibinfo{editor}{M.~Piazzesi}, \& \bibinfo{editor}{L.~Samuelson} (Eds.), {\it
  \bibinfo{booktitle}{Advances in Economics and Econometrics: Eleventh World
  Congress}\/} (p. \bibinfo{pages}{176–214}).
\newblock \bibinfo{publisher}{Cambridge University Press}
  volume~\bibinfo{volume}{1} of {\it \bibinfo{series}{Econometric Society
  Monographs}\/}.
%Type = Article
\bibitem[{Pittel(1989)}]{pittel1989}
\bibinfo{author}{Pittel, B.} (\bibinfo{year}{1989}).
\newblock The Average Number of Stable Matchings.
\newblock {\it \bibinfo{journal}{SIAM Journal on Discrete Mathematics}\/},
  {\it \bibinfo{volume}{2}\/}, \bibinfo{pages}{530--549}.
%Type = Article
\bibitem[{Roth \& Peranson(1999)}]{roth1999}
\bibinfo{author}{Roth, A.~E.}, \& \bibinfo{author}{Peranson, E.}
  (\bibinfo{year}{1999}).
\newblock The Redesign of the Matching Market for American Physicians: Some
  Engineering Aspects of Economic Design.
\newblock {\it \bibinfo{journal}{American Economic Review}\/},  {\it
  \bibinfo{volume}{89}\/}, \bibinfo{pages}{748--780}.
%Type = Incollection
\bibitem[{S{\"o}nmez \& {\"U}nver(2011)}]{sonmezunver2011}
\bibinfo{author}{S{\"o}nmez, T.}, \& \bibinfo{author}{{\"U}nver, M.~U.}
  (\bibinfo{year}{2011}).
\newblock Matching, Allocation, and Exchange of Discrete Resources.
\newblock In {\it \bibinfo{booktitle}{Handbook of Social Economics}\/} (pp.
  \bibinfo{pages}{781--852}).
\newblock \bibinfo{publisher}{Elsevier} volume~\bibinfo{volume}{1}.
%Type = Article
\bibitem[{Wu \& Zhong(2014)}]{wu2014}
\bibinfo{author}{Wu, B.}, \& \bibinfo{author}{Zhong, X.}
  (\bibinfo{year}{2014}).
\newblock Matching Mechanisms and Matching Quality: Evidence from a Top
  University in China.
\newblock {\it \bibinfo{journal}{Games and Economic Behavior}\/},  {\it
  \bibinfo{volume}{84}\/}, \bibinfo{pages}{196--215}.

\end{thebibliography}

\end{document}